\documentclass{elsart}
\usepackage{graphicx}
\usepackage{amsmath}

\def\la{\mathrel{\mathpalette\fun <}}
\def\ga{\mathrel{\mathpalette\fun >}}
\def\fun#1#2{\lower3.6pt\vbox{\baselineskip0pt\lineskip.9pt
  \ialign{$\mathsurround=0pt#1\hfil##\hfil$\crcr#2\crcr\sim\crcr}}}

\begin{document}
\begin{frontmatter}

\title{CRPropa: A Numerical Tool for the Propagation of UHE Cosmic Rays, $\boldsymbol{\gamma}-$rays and Neutrinos}

\author{Eric~Armengaud$^{a,b,c}$, G\"unter~Sigl$^{a,b}$, 
Tristan~Beau$^a$, Francesco~Miniati$^{d}$}

\address{$^a$APC, B\^atiment Condorcet, 10 rue Alice Domon et L\'eonie Duquet, 75205 Paris Cedex 13, France - UMR 7164 (CNRS, Universit\'e Paris Diderot Paris 7, CEA, Observatoire de Paris)}

\address{$^b$ GReCO, Institut d'Astrophysique de Paris, C.N.R.S.,
98 bis boulevard Arago, F-75014 Paris, France}

\address{$^c$ CEA, DSM, DAPNIA, Centre d'Etudes de Saclay, F-91191 Gif-sur-Yvette Cedex, France}

\address{$^d$ Physics Department, ETH Z\"urich, 8093 Z\"urich, 
Switzerland}

\begin{abstract}
To understand the origin of ultra-high energy cosmic rays (UHECRs,
defined to be above $10^{18}\,$eV), it is required to model in a realistic way their propagation in the Universe. UHECRs can interact
with low energy radio, microwave, infrared and optical photons to produce
electron/positron pairs or pions. The latter decay and give rise to neutrinos
and electromagnetic cascades extending down to MeV energies. In addition, deflections in cosmic magnetic fields can influence the spectrum and sky 
distribution of primary
cosmic rays and, due to the increased propagation path length, the
secondary neutrino and $\gamma-$ray fluxes.
Neutrino, $\gamma-$ray, cosmic ray physics and extra-galactic magnetic fields
are, therefore, strongly linked subjects and should be considered together
in order to extract maximal information from existing and future data, like the
one expected from the Auger Observatory. For that purpose,
we have developed CRPropa, a publicly-available numerical package
which takes into account interactions and
deflections of primary UHECRs as well as propagation of secondary electromagnetic
cascades and neutrinos. CRPropa allows to compute the observable
properties of UHECRs and their secondaries in a variety of models for
the sources and propagation of these particles. Here we present
physical processes taken into account as well as benchmark examples; a
detailed documentation of the code can be found on our web site.
\end{abstract}

\end{frontmatter}

\section{Introduction}
Astroparticle physics is currently experimentally driven and involves
many different existing or planned projects ranging from UHECR
observatories such as the Pierre Auger Observatory~\cite{auger}, to neutrino
telescopes~\cite{nu_review}, as well as ground and space based $\gamma-$ray
detectors operating at TeV and GeV energies,
respectively~\cite{gammarev}. It is clear that GeV-TeV $\gamma-$ray
and neutrino astronomy will prove an invaluable tool to unveil the
sources, and probe into the mechanism, of UHECRs.  Even if a putative
source were to produce exclusively UHECRs, photo-pion~\cite{gzk} and
pair production by protons on the cosmic microwave background (CMB)
would lead to guaranteed secondary photon and neutrino fluxes that
could be detectable. Furthermore, spectra, power and sky distributions
of both primary UHECRs and secondary $\gamma-$rays and neutrinos
depend on the poorly known large scale cosmic magnetic fields.

It is, therefore, desirable to have a numerical tool that can treat
the interface between UHECR, $\gamma-$ray and neutrino astrophysics, and
large scale magnetic fields. To this end,
we have recently merged our Monte Carlo code for simulating
three dimensional propagation of UHECRs in a structured, magnetized 
Universe~\cite{Sigl:2004yk}
with a one-dimensional transport code that solves electromagnetic (EM)
cascades and neutrino propagation~\cite{Lee:1996fp}. 
We discuss the limitations due to the one-dimensional approximation
and implement a procedure to test the resulting uncertainty on the EM cascade
on the observable fluxes.
With the present paper, we release a public version of this code which
we hope will be useful for the cosmic ray, $\gamma-$ray and neutrino
communities.

In the following, we present the relevant interactions and propagation
phenomena taken into account, and the propagation algorithms
applied in CRPropa. We also present a few examples of how to use
the code in practice. The numerical package
and its detailed documentation are available for downloading
on the CRPropa website, {\tt http://apcauger.in2p3.fr/CRPropa}.

We use natural units, $c=\hbar=1$ throughout this paper.

\section{Propagation algorithms}

UHECRs are injected at specified sources, and propagated
step-by-step in either a one- or a three-dimensional environment. The
trajectories are regularly sampled, or recorded only at specific
locations (e.g. at a given distance from a source, or at an
``observer'' point). Each propagation step consists of integrating the
Lorentz equations, and computing the interactions and possibly the
secondaries generated by those interactions.

In the 3-dimensionnal case, a ``simulation box'' is defined and
periodic boundary conditions are assumed.

When deflections are taken into account, cosmological redshifts cannot 
be computed,
because the propagation time until the particle reaches the observer
is not known before hand. Therefore, redshift evolution is only
accounted for in the 1D version of the package. The concordance
cosmology is used for which,
assuming a flat Universe, 
the Hubble rate $H(z)$ at redshift $z$
in the matter dominated regime, $z\la 10^3$, is given by
\begin{equation}\label{cosmo}
  H(z)= H_0
  \left[\Omega_{\rm m}(1+z)^3+\Omega_{\Lambda}\right]^{1/2}\,.
\end{equation}
The parameters $\Omega_{\rm m}$ and $\Omega_{\Lambda}$ can be freely
chosen, their standard values being $\Omega_{\rm m}=0.3$,
$\Omega_{\Lambda}=0.7$, and $H_0=h_0\,100~{\rm km}~{\rm s}^{-1}~{\rm
Mpc}^{-1}$ with $h_0=0.72$.

\begin{figure}
\begin{center}
\includegraphics[width=0.8\textwidth]{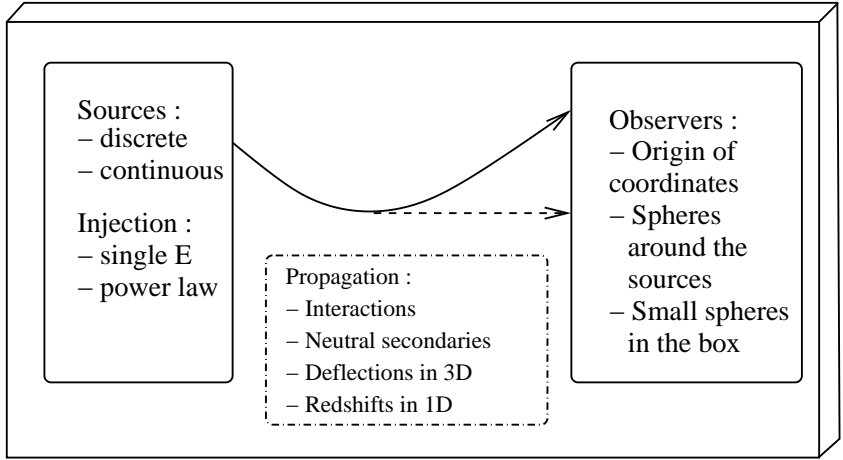}
\caption{\label{crp_graph} Principle of the propagation
algorithm. This scheme applies to all configurations.}
\end{center}
\end{figure}

The general principle of the simulations is shown in
Fig.~\ref{crp_graph}.

\subsection{Nucleon Interactions} 

The most famous interaction of nucleons with the low-energy photon
backgrounds is pion production, which generates the GZK feature. In
order to handle pion production, we use the event generator
SOPHIA~\cite{sophia}, that has been explicitely designed to study
this phenomenon and that uses the particle production
cross-sections measured in accelerators. We have also augmented the SOPHIA 
package for interactions with a low energy extragalactic background
light (EBL) with a general energy distribution. SOPHIA allows to
determine the distribution of the stables particles generated by an
interaction with a low-energy photon.

Pair production by protons (PPP) on the CMB, also known as 
Bethe-Heitler process, is taken into account as a continuous energy loss whose 
rate we evaluate following the expressions in 
Refs.~\cite{Blumenthal:1970nn,chodorowski}.
For the spectrum of the pairs we exploit the fact that Bethe-Heitler
and triplet pair production, $e\gamma_b\to ee^+e^-$, are analogous
electromagnetic processes, their cross sections and inelasticities
converging for relativistic pairs. Fig.~2 of Ref.~\cite{mastichiadis}
then shows that the spectrum of electron-positron pairs (heretoafter
simply referred to as electrons) generated by a proton of energy $E$
can be approximated by a power-law energy distribution $dn/dE_e\propto
E_e^{-7/4}$. Kinematics implies that this power law holds for 
$E_{\rm min}\leq E_e\leq E_{\rm PPP}$, where the minimal
and maximal energies are given by~\cite{Lee:1996fp}
\begin{eqnarray}
  E_{\rm PPP}&\simeq&\frac{4E^2\varepsilon}{4E\varepsilon+m_p^2}
  \simeq\frac{4.5\times10^{15}\left(\frac{E}{{\rm EeV}}\right)^2
  \left(\frac{\varepsilon}{{\rm meV}}\right)\,{\rm eV}}
  {4.6\times10^{-3}\left(\frac{E}{{\rm EeV}}\right)
  \left(\frac{\varepsilon}{{\rm meV}}\right)+1}\nonumber\\
  E_{\rm min}&\simeq&\frac{m_e^2}{8\varepsilon}\simeq3.3\times10^{13}\,
  \left(\frac{\varepsilon}{{\rm meV}}\right)^{-1}\,{\rm eV}
  \,.\label{E_ppp}
\end{eqnarray}
In Eq.~(\ref{E_ppp}), $m_p$ and $m_e$ are the proton and electron masses,
respectively, $\varepsilon$ is the low energy target photon energy, and
the approximation for $E_{\rm min}$ holds for
$m_e m_p\la 4E\varepsilon\la m_p^2$.
The average electron energy is then $\overline{E_e}=
\int^{E_{\rm PPP}}_{E_{\rm min}}dE_e E_e E_e^{-7/4} /
\int^{E_{\rm PPP}}_{E_{\rm min}}dE_e E_e^{-7/4}
\simeq3\,E_{\rm min}^{3/4}E_{\rm PPP}^{1/4}$ which is indeed much smaller
than the primary proton energy $E$. From Eq.~(\ref{E_ppp}), the
inelasticity $K\equiv\overline{E_e}/E$,
whose precise energy dependence can be found in Ref.~\cite{chodorowski}, for
$m_e m_p\la4E\varepsilon\la m_p^2$ can thus be approximated by
\begin{eqnarray}
  K(E\varepsilon)&\sim&\frac{3}{2^{7/4}}
  \frac{m_e^{3/2}}{\left(E\varepsilon\,m_p\right)^{1/2}}\label{K}\\
  &\simeq&3.4\times10^{-4}\left(\frac{E}{{\rm EeV}}\right)^{-1/2}
  \left(\frac{\varepsilon}{{\rm meV}}\right)^{-1/2}\,,\nonumber
\end{eqnarray}
This is consistent with Figs.~1 and~2 in Ref.~\cite{Mastichiadis:2005nj}.
For our purposes,
we are not sensitive to the lower kinematic limit since the total energy
produced $\propto\int^{E_{\rm PPP}}_{E_{\rm min}}dE_e E_e E_e^{-7/4}
\simeq4E_{\rm PPP}^{1/4}$ is insensitive to $E_{\rm min}$ as long as
$E_{\rm min}\ll E_{\rm PPP}$, but rather is dominated by the highest energies.
As a consequence, the total proton energy loss rate due to
pair production is dominated by the highest energy electrons close to
$E_{PPP}$. However, because the production cross section of these
highest energy electrons is much smaller than the one
for the more numerous lower energy electrons, the average inelasticity
Eq.~(\ref{K}) is nevertheless small, below $10^{-3}$ everywhere above
the pair production threshold. The spectrum
and maximal energy of the pairs will be important for
the synchrotron spectrum emitted by these electrons in an EGMF of strength $B$ 
which peaks at
$\simeq6.8\times10^{11}\,(E_e/10^{19}\,{\rm eV})^2(B/0.1\,\mu{\rm G})\,$eV.

Nucleons can be followed down to $10^{17}\,$eV with CRPropa, below which 
interactions become negligible.

\subsection{Secondary Electromagnetic Cascades and Neutrinos}

The secondary neutrinos from pion production of nucleons are propagated
in straight lines assuming no energy losses except redshift effects.

All the EM products of these interactions are evolved using
an EM cascade code based on Ref.~\cite{Lee:1996fp}. The photons and
pairs are followed until either their energy drops below 100 MeV
or they reach an observer. All relevant interactions
with a background photon $\gamma_b$ are taken into account, namely single pair
production (PP), $\gamma\gamma_b\to e^+e^-$, double pair production (DPP),
$\gamma\gamma_b\to e^+e^-e^+e^-$, inverse Compton scattering (ICS),
$e\gamma_b\to e\gamma$, and triplet pair production (TPP), $e\gamma_b\to ee^+e^-$
(see also Ref.~\cite{bs} for a detailed discussion of implemented interactions).
In addition, synchrotron losses of electrons in the (in 
general) inhomogeneous EGMF are
taken into account and the resulting lower energy synchrotron photons are
also followed in the subsequent EM cascade.

This module has been applied to EM cascades from discrete magnetized
proton sources in galaxy clusters in Ref.~\cite{Armengaud:2005cr}.
The EM cascades that are followed with the current version of CRPropa
are propagated in straight lines, even in the case of 3-dimensionnal
simulations for UHECRs:
Every time a primary hadron interacts and initiates an EM
cascade, it is assumed that the secondaries propagate
along straight lines and it is checked whether the line of sight crosses
the observer. If this is the case, the EM cascade module is called with the 
corresponding propagation distance and the projected magnetic field profile.
Electrons in the EM cascade can of course be deflected in the
EGMF, and we discuss here the validity of this one-dimensionnal approximation.

\begin{figure}
\begin{center}
\includegraphics[width=0.8\textwidth]{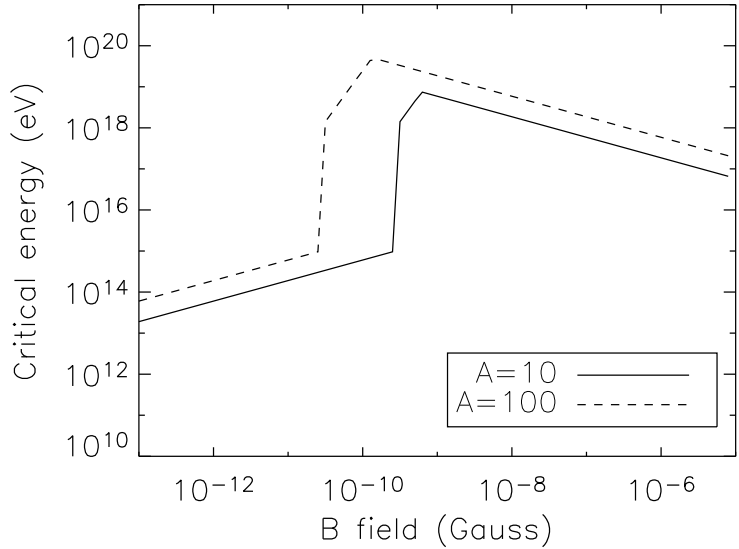}
\caption{\label{fig:ecrit} The critical energy $E_c$, below which the $e^{+/-}$
are deflected before cascading to lower energies, as a function of the
order of magnitude of the magnetic field. $E_c$ is obtained with the
parameterization of various timescales given in the text, for $A=10$ (solid
line) and $A=100$ (dashed line). This corresponds to cutting all pairs
being deflected by more than $\sim6^\circ$ or $0.6^\circ$, respectively.
Note that the jump around $\simeq3\times10^{-10}\,$G and $\simeq2\times10^{-11}\,$G, respectively, is due to the transition
from ICS to synchrotron emission (at large fields) as dominant energy
loss.}
\end{center}
\end{figure}

In a magnetic field of strength $B$ the synchrotron cooling time
for an electron of energy $E_e$ is given by
\begin{eqnarray}
  t_{\rm synch}&=&\frac{E_e}{dE_e/dt}=\frac{6\pi m_e^2}{\sigma_T E_e B^2}
  \label{synchro}\\
  &\simeq&3.84\,{\rm kpc}\,\left(\frac{E_e}{10^{15}\,{\rm eV}}\right)^{-1}
  \left(\frac{B}{\mu\,{\rm G}}\right)^{-2}\,,\nonumber
\end{eqnarray}
where $\sigma_T = 8 \pi \alpha^2 / 3 m_e^2$ is the Thomson cross
section, with $\alpha$ the fine structure constant.
At high energies, in the Klein-Nishina regime the inverse Compton energy loss length is 
roughly~\cite{bs}
\begin{equation}
  t_{\rm IC}\la 400\,{\rm pc}\,\left(\frac{E_e}{10^{15}\,{\rm eV}}\right)
  \quad\mbox{for}\,E_e\ga 10^{15}\,{\rm eV}\,.\label{ic}
\end{equation}
At energies $E_e\ga 10^{18}\,$eV in Eq.~(\ref{ic}) the
energy loss length is between a factor $\sim30$ and a few hundred smaller
than the numerical value in Eq.~(\ref{ic}) due to contributions from the
universal radio background. For a conservatively large $t_{\rm IC}$ at these energies we use an interpolation of Fig.~12 in~\cite{bs} for the
conservatively low radio background estimate.
For $E_e\la 10^{15}\,$eV,
ICS on the CMB is in the Thomson regime, with an interaction length $\lambda_{\rm IC} \sim
1/\sigma_T n_{\rm CMB} \sim 1.2$ kpc, with $n_{\rm CMB}$ the CMB
photon density. The energy lost by an electron
at each interaction is $\delta E_e \sim 4 \epsilon E_e^2/ 3 m_e^2$,
where $\epsilon$ is a typical CMB photon energy. As a consequence, the
energy loss length at energies below $\sim 10^{15}$ eV is:
\begin{equation}
t_{\rm IC} \simeq \frac{3 \lambda_{\rm IC} m_e^2}{4 \epsilon E_e} \sim
400\,{\rm pc}\,\left( \frac{10^{15}\,\rm{eV}}{E_e}\right)\,. \label{ict}
\end{equation}
These length scales as well as the maximal propagation distance must 
be compared with the Larmor radius
\begin{equation}
  r_L=\frac{E_e}{eB}\simeq1.08\,{\rm pc}\,
  \left(\frac{E_e}{10^{15}\,{\rm eV}}\right)
  \left(\frac{B}{\mu\,{\rm G}}\right)^{-1}\,.\label{Larmor}
\end{equation}
In order for a one-dimensional treatment of EM cascades
to be a good approximation, the Larmor radius has to be much larger than either
the total propagation length, the IC or the synchrotron
loss lengths. For a given magnetic field, the condition $r_L >
A\times\rm{min}(t_{\rm synch}, t_{\rm IC})$ results in a condition $E_e >
E_c(B,A)$, corresponding to deflections of the pairs by
$\la(10/A)\times 6^{\circ}$. This estimate of the deflection angle is
however conservatively large: in realistic situations, magnetic fields
are inhomogeneous with many reversals along the line of sight,
and the actual deflection angle will be smaller provided the magnetic
field coherence length is smaller than the energy loss length.
The dependence of the ``critical''
energy $E_c$ on $B$ and $A$ is shown in Fig.~\ref{fig:ecrit}:
For $A=10$, corresponding to deflection by less than $\sim6^\circ$,
$E_c$ is determined by the competition between deflections and ICS
for $B\la 300$ pG, or between deflection and synchrotron emission
for $B\ga 300$ pG. For $A=100$, corresponding to deflection by less than 
$\sim0.6^\circ$, the transition between ICS and synchrotron
emission as dominant losses to be compared with deflection occurs
at $B\simeq20$ pG.

It turns out that whenever $E_c(B,A)\la10^{15}\,$eV, the $\gamma-$ray
flux from deflected pairs is sub-dominant. This is simply due to the
fact that the pair flux at energies $E_e\la10^{15}\,$eV is suppressed
compared to the $\gamma-$ray flux which has a much larger interaction length
and piles up below the pair production threshold. The $\gamma-$ray
flux from deflected pairs can only be important if $E_c(B,A)\gg10^{15}\,$eV
which, from Fig.~\ref{fig:ecrit}, requires that synchrotron emission dominates
the losses of the deflected pairs. In this case, a significant fraction
of the energy flux going into pairs is deflected more than
$\sim(10/A)\times 6^{\circ}$, thus modifying
the $\gamma-$ray point flux at energies
\begin{equation}\label{eq:e_synch}
E_{\gamma} \la 2.2\times 10^{8} \,{\rm eV} \left(\frac{E_c(B,A)}{\rm
EeV} \right)^2 \left( \frac{B}{\rm nG}\right)\,.
\end{equation}
In the following we will confirm these expectations with numerical
simulations.

\begin{figure}
\begin{center}
\includegraphics[width=0.8\textwidth]{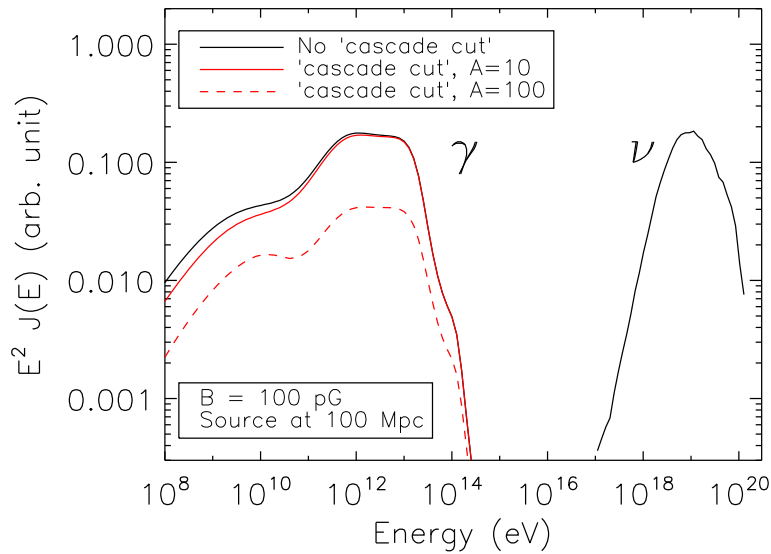}
\caption{\label{fig:cascade_cut} Flux of secondary neutrinos (all
flavors are added) and $\gamma$-rays from a single source of UHE
protons with injection spectrum $\propto E^{-2}$ up to
$5\times 10^{20}$ eV, computed assuming a straight line propagation
for the protons but taking into
account the influence of a 100 pG magnetic field on the EM
cascades. The red line is the flux computed by ``cutting'' the
local $e^{+/-}$ flux below $E_c(B,A)$ (see text), for $A=10$
(continuous line) and $A=100$ (dashed line). This corresponds to
pair deflections of $6^{\circ}$ or $0.6^{\circ}$, respectively.}
\end{center}
\end{figure}

Within CRPropa, the parameter $A$ can be chosen by the user,
and the local contribution of electrons with energy $E_e
< E_c(B,A)$ to the $\gamma$-ray flux can be switched on or off: This
allows to estimate the uncertainty in the $\gamma$-ray flux arriving
within a certain angle $\sim(10/A)\times 6^{\circ}$ from a point
source due to the 1D approximation. An example is shown in 
Fig.~\ref{fig:cascade_cut}, where the computed $\gamma$-ray fluxes
from a single proton source located at 100 Mpc from the Earth in
a uniform magnetic field of amplitude 100 pG are compared with and
without cutting the charged component of the EM cascade deflected by more
than $6^\circ$ and $0.6^\circ$, respectively.

For the flux arriving within $6^\circ$ in Fig.~\ref{fig:cascade_cut}, 
$E_c\simeq3\times10^{14}\,$eV,
see Fig.~\ref{fig:ecrit}, and indeed a discernible but still modest,
$\sim30\%$, modification appears only for $E_{\gamma} \la 0.1$ TeV, where
the photon energy flux becomes comparable to the pair energy
flux around $E_c$.

For the flux arriving within $0.6^\circ$ in Fig.~\ref{fig:cascade_cut}, 
$E_c\simeq2\times10^{19}\,$eV, see Fig.~\ref{fig:ecrit}, and by the above
argument and Eq.~(\ref{eq:e_synch}) we expect the photon flux to be significantly
modified below $\sim10\,$TeV. Indeed, at these energies the flux is
reduced by a factor $\sim5$.

In case of comparatively strong magnetic fields of order $\mu$G,
typical in galaxy clusters, $E_c\la10^{18}\,$eV, and $\gamma-$ray
point fluxes arriving within $\sim0.6^\circ$ should only be modified 
significantly for $E_\gamma\la100\,$GeV. Also
note that for the production of secondaries inside a magnetized
region where the parent UHECR particles are isotropically
distributed, the full three dimensional treatment of the EM
cascade is not necessary
because for any $e^-$ that is deflected away from the line of sight
there is always another $e^-$ that is deflected into the line of sight.
In realistic situations, the magnetic fields are highly
structured with typical amplitudes of $\sim \mu G$ in the clusters,
and $\la 10$ pG in the voids. The above discussion shows that in
all these cases CRPropa can estimate the minimal $\gamma-$ray flux
arriving within an angle $(10/A)\times 6^{\circ}$ from the source.

\subsection{Background Photon Spectra and their Evolution}

\begin{figure}
\begin{center}
\includegraphics[width=0.8\textwidth]{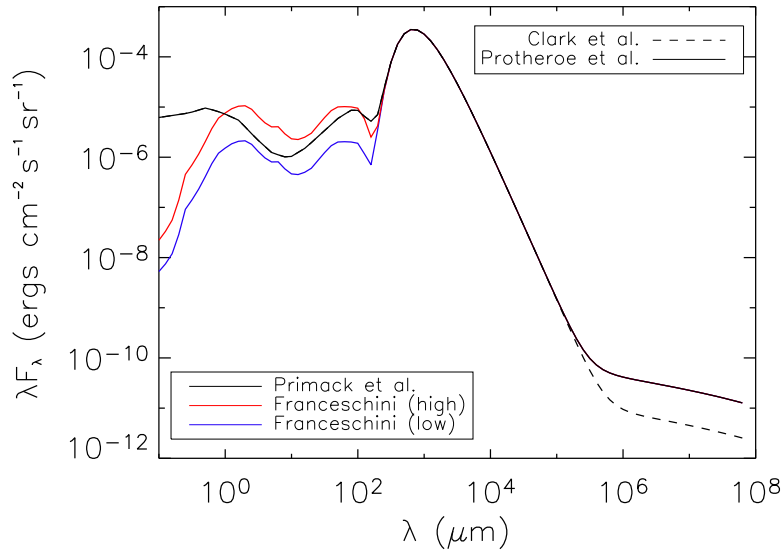}
\caption{\label{fig1}Models implemented for the low energy photon background at
zero redshift. The IRB consists basically of a peak in the far infrared around
100$\mu\,$m dominated by dust and a peak in the near infrared dominated by stars.}
\end{center}
\end{figure}

Fig.~\ref{fig1} shows the EBL energy distributions that have been
implemented. The most important is the CMB.
For the infrared background (IRB) we implemented three
distributions, a low and a high version of Franceschini 
et al.~\cite{Franceschini:2001yg}
which differ roughly by a factor 5, as well as the one by Primack
et al.~\cite{Primack:2005rf}. The low Franceschini et al. and the
Primack et al. backgrounds are consistent with recent upper limits
from blazar observations in TeV $\gamma-$rays by HESS~\cite{Aharonian:2005gh}.
For a recent review of the IRB see for example Ref.~\cite{Lagache:2005sw}.

The IRB has a significant influence on EM cascades only
around the threshold for pair production, i.e. between a few Tev and
$\simeq100\,$TeV. At higher energies, the $\gamma-$ray flux is suppressed by 
interactions with the CMB and, above $\simeq10^{19}\,$eV, by interactions with 
with the radio background. At energies below
$\sim\,$TeV, the Universe acts as a calorimeter and the total photon
flux is proportional to the total EM energy injected above $\sim\,$PeV
with a rather universal shape~\cite{Coppi:1996ze}.

Although its photon number
density $\simeq2\,{\rm cm}^{-3}$ is a factor $\simeq200$ smaller than for
the CMB, below the GZK-cutoff and above $\sim10^{17}\,$eV the IRB
can significantly reduce the nucleon mean free path for pion production.
This can be important
for secondary photon and neutrino~\cite{Stanev:2004kz,Bugaev:2004xt}
production, especially for a steep primary injection spectrum and/or
strong redshift evolution.

For the universal radio background (URB) we use a weak and a strong version
based on Ref.~\cite{Protheroe:1996si} and on observations~\cite{obs_radio}.
The URB is mostly important for EM cascades above $\sim10^{18}\,$eV where
it can inhibit cascade development due to the resulting small pair production
lengths, especially for fast synchrotron losses of electrons in the
presence of strong magnetic fields.

Since URB photons can give rise to pion
production only above a few times $10^{22}\,$eV, where the interaction rate
is essentially proportional to the total EBL photon density which is dominated
by the CMB by a factor $\sim10^3$, see Fig.~\ref{fig1}, the URB is negligible
for pion production. The same applies to pair production by protons.

\begin{figure}
\begin{center}
\includegraphics[width=0.8\textwidth]{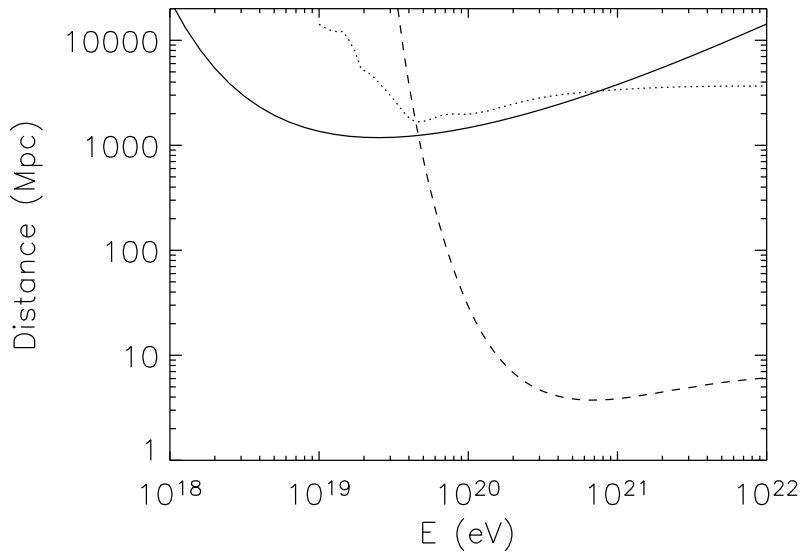}
\caption{\label{fig2} Proton energy loss length for pair production on
the CMB (continuous line), interaction length for pion production on
the CMB (dashed line) and on the Primack et al. IRB (dotted line) at
$z=0$. The irregularities in the dashed curve are due to the
piecewise power law fits of the Primack et al. IRB.}
\end{center}
\end{figure}

\begin{figure}
\begin{center}
\includegraphics[width=0.8\textwidth]{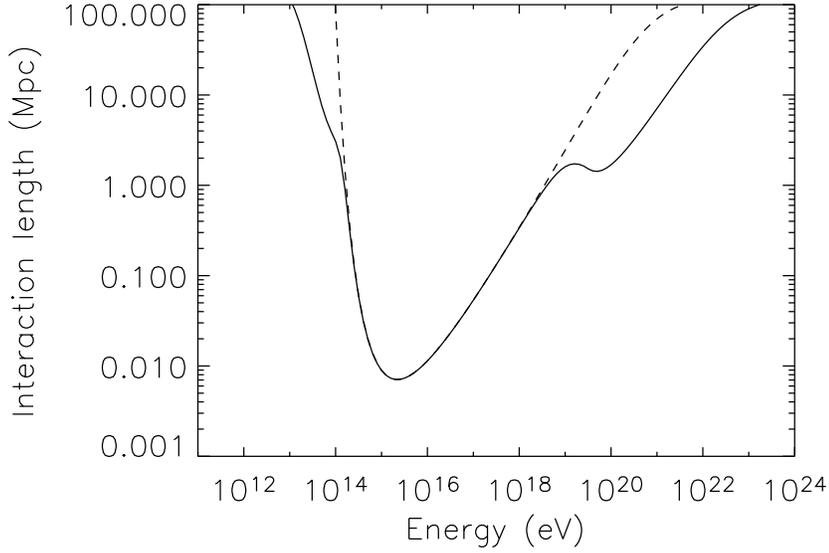}
\caption{\label{fig3}Photon interaction length at $z=0$ on the EBL
consisting of the CMB, the Primack
et al. IRB, and the strong URB version. Dotted line: Interaction
length in the CMB only at $z=0$.}
\end{center}
\end{figure}

Figs.~\ref{fig2} and~\ref{fig3} show interaction and energy loss lengths
for protons and interaction lengths of photons, respectively, and their
dependence on EBL models at zero redshift. This demonstrates that the
IRB becomes important
for pion production by protons below the GZK cutoff and for pair production
by photons below the threshold in the CMB at $\sim10^{14}\,$eV. It also shows
that the URB tends to dominate pair production by photons above $\sim10^{19}\,$eV.

The redshift evolution of the cosmic microwave background
(CMB) is trivial. The redshift evolution of the radio and infrared
distributions is more complicated: Ultra-relativistic particles of energy
$E$ injected at redshift $z^\prime$ with a rate per energy and comoving volume
$\Phi(E,z^\prime)$ result in a {\it physical} number density per energy
at redshift $z$ given by
\begin{eqnarray}
  n(E,z)&=&(1+z)^3\int_z^\infty
  dz^\prime\frac{4\pi\Phi\left[E_i(E,z,z^\prime),z^\prime\right]}
           {(1+z^\prime)H(z^\prime)}\nonumber\\
           &&\hskip2cm\times\frac{dE_i}{dE}(E,z,z^\prime)\,,\label{bkg1}
\end{eqnarray}
where it is assumed that the particle looses energy continuously such that
its injection energy can be computed analytically, $E_i(E,z,z^\prime)$.
Interactions of the low energy EBL photons, whose differential number densities
we will denote by $n_b(\varepsilon,z)$ in the following to distinguish
from the high energy particles, can safely be neglected after
recombination, $z\la10^3$, such that $E_i(E,z,z^\prime)=(1+z^\prime)E/(1+z)$.
Eq.~(\ref{bkg1}) then simplifies to
\begin{equation}
  n_b(\varepsilon,z)=(1+z)^2\int_z^\infty
  dz^\prime\frac{4\pi\Phi\left[(1+z^\prime)\varepsilon/(1+z),z^\prime\right]}
           {H(z^\prime)}\,,\label{bkg2}
\end{equation}
By using $|dt/dz|=[(1+z)H(z)]^{-1}$, one can see easily that the total energy
density per comoving volume redshifts as $\int d\varepsilon\, \varepsilon\,
n_b(\varepsilon,z)/(1+z)^3=(1+z)
\int dt\,d\varepsilon_i\,\Phi(\varepsilon_i,z^\prime)/(1+z^\prime)$, as it should be.

For the URB we implemented a nontrivial redshift evolution in the cascade module,
as this can be relevant for EM cascade development. We assume that
$\Phi_{\rm URB}(\varepsilon,z)=\phi_{\rm URB}(\varepsilon)g_{\rm URB}(z)$
factorizes into an energy dependence $\phi_{\rm URB}(\varepsilon)$ motivated
by the observations~\cite{obs_radio} and theoretical estimates~\cite{Protheroe:1996si}
and a redshift dependence given by
\begin{equation}
  g_{\rm URB}(z)=10^{1.18z-0.28z^2}\,,
\end{equation}
as in Ref.~\cite{Lee:1996fp}.

For the Primack et al. IRB~\cite{Primack:2005rf} we use for simplicity
the differential photon energy distribution evolution
\begin{equation}
  n_b(\varepsilon,z)=
  \left\{\begin{array}{ll}
  (1+z)^2n_b\left(\frac{\varepsilon}{1+z},z=0\right)\,
  & \mbox{for $z\leq z_b$}\,,\\
0 & \mbox{otherwise}
\end{array}\right\}\label{trivial_evol}
\end{equation}
which corresponds to instantaneous creation of the background at redshift
$z_b$ with $\Phi(\varepsilon,z^\prime)=H(z_b)n_b[\varepsilon/(1+z_b),z=0]
\delta(z^\prime-z_b)/(4\pi)$ in Eq.~(\ref{bkg2}). It strictly applies to
the CMB which was effectively produced at decoupling, $z_b\sim1100$.
For the IRB we assume $z_b=5$. Interaction
lengths $l(E,z)$ and, in case of continuous energy loss processes such as PPP,
energy loss rates $b(E,z)\equiv dE/dt$ then follow simple scaling relations
in redshift~\cite{Bugaev:2004xt},
\begin{eqnarray}
  l(E,z)^{-1}&=&(1+z)^3l\left[(1+z)E,z=0\right]^{-1}\nonumber\\
  b(E,z)&=&(1+z)^2b\left[(1+z)E,z=0\right]\,.\label{scaling}
\end{eqnarray}
This simplifies implementation in SOPHIA.

\subsection{Distributions and Properties of Sources}
Both single sources and realizations of both discrete or continuous source
distributions can be used in CRPropa. In the latter case, the distributions can
be selected, for example, to follow the baryon density from a large
scale structure simulation box, and are periodically repeated.

The UHECR particles are injected isotropically around the sources with
a monochromatic or a power-law energy distribution between
a minimal and a maximal energy,
$E_{\rm min}$ and $E_{\rm max}$, respectively:

$$ \frac{dN}{dE_{\rm inj}} \propto E_{\rm inj}^{-\alpha}
\qquad E_{\rm min} \leq E_{\rm inj} \leq E_{\rm max}$$

For each trajectory reaching the observer and being registered, the
source identity $i$ is also registered. This allows to apply a re-weighting procedure
on the recorded ``events'', in order to vary individual source properties such
as their injection power law index $\alpha_i$
or luminosity $Q_i$. For example, it is most efficient in terms
of CPU time
to inject the UHECRs with a spectral index $\alpha_0 = 1$ at the
sources, that is with a uniform distribution in the logarithm of the
energy. By re-weighting each recorded event by a factor
$ w \propto Q_i E_{\rm inj}^{\alpha_i-1} $, the source $i$ would
contribute with a power $Q_i$ and an effective injection power law index
$\alpha_i$ in all observables constructed from the weighted trajectory
sample.

\section{Large Scale Structure and Magnetic Fields}

The strength and distribution of the EGMF is currently poorly known and their
impact on UHECR are hard to quantify, as demonstrated by the different results
in Refs.~\cite{Sigl:2004yk,dolag}. See also Ref.~\cite{Sigl:2004gi} 
for a discussion of
these differences and Ref.~\cite{bo_review} for a review on EGMF. We note that
there are recent observational hints of EGMF as strong as $\sim0.1\,\mu$G on
scales as extended as superclusters~\cite{Xu:2005rb}, as well as
theoretical motivations for such fields~\cite{Medvedev:2005ep}.

Enhanced magnetic fields around large scale structures such as
galaxy clusters together with associated larger EBL densities can lead
to increased production of $\gamma-$rays and neutrinos.

\begin{figure}
\begin{center}
\includegraphics[width=0.7\textwidth]{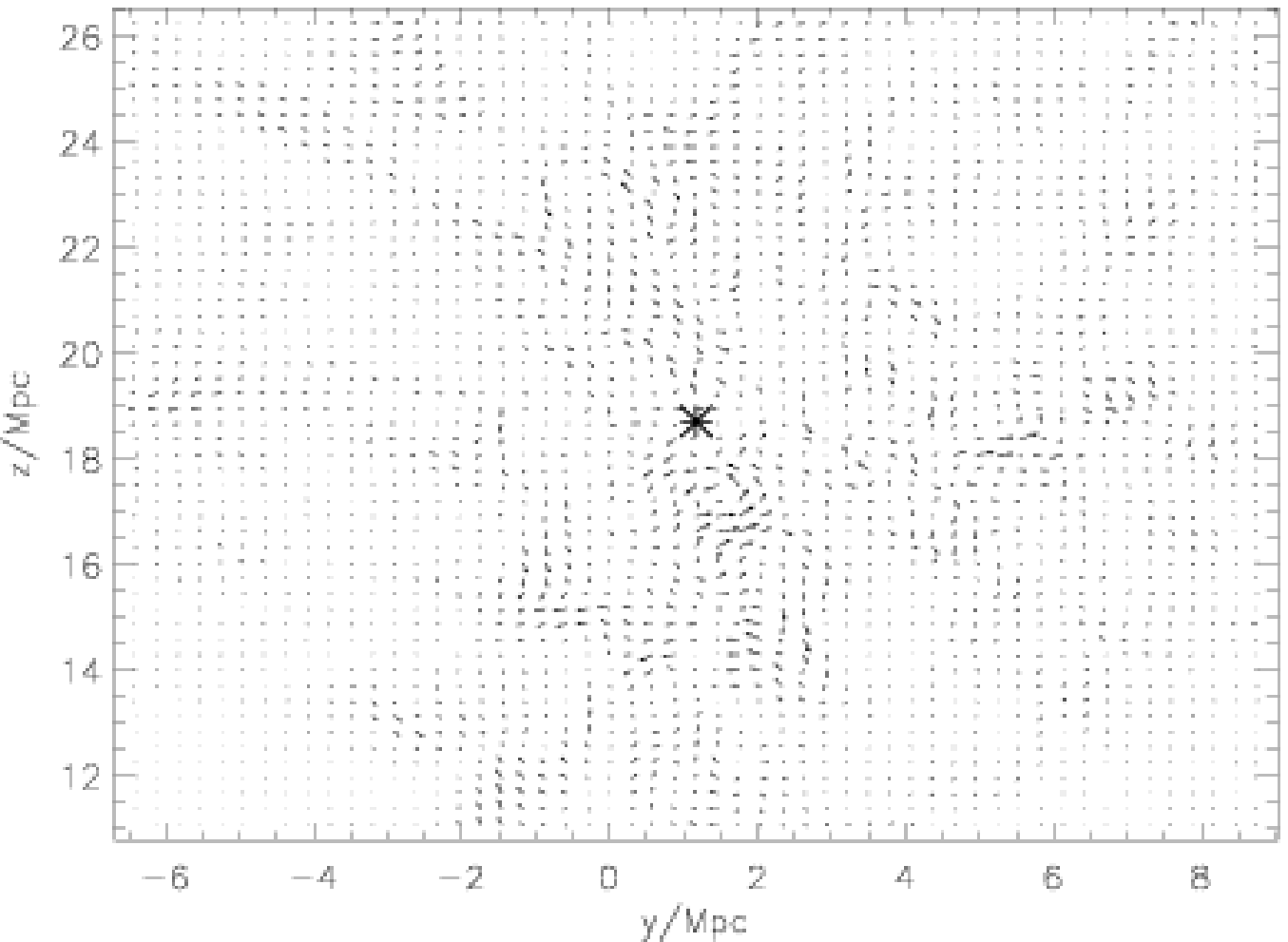}
\includegraphics[width=0.6\textwidth]{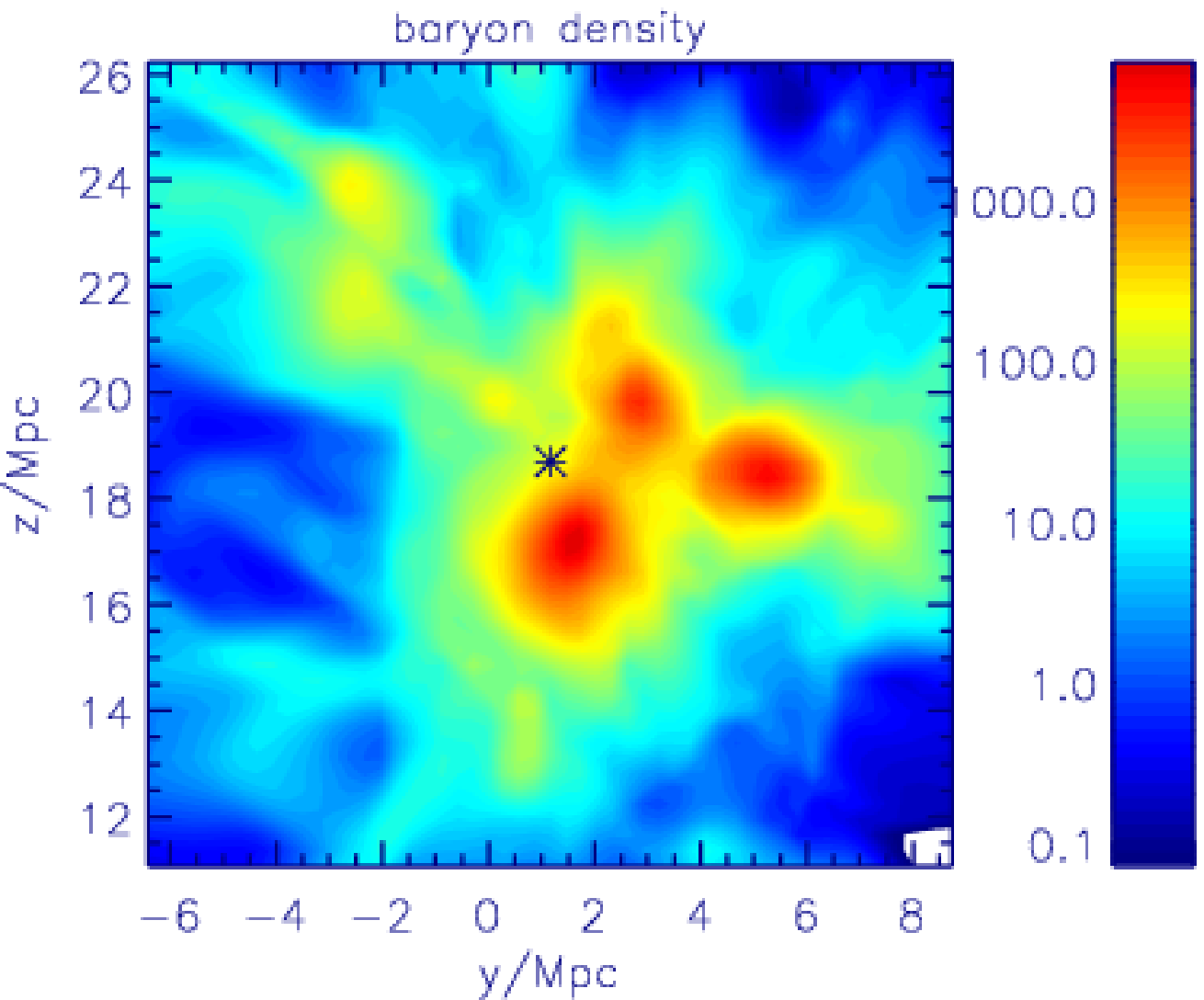}
\caption{\label{fig4}A 2D cross section through the relative size and
polarization of the EGMF in linear scaling, 
(top panel) and the relative
baryon density in logarithmic scaling (bottom panel) 
in the environment of a galaxy cluster from the simulations from
Ref.~\cite{ryu,miniati}.}
\end{center}
\end{figure}

The EGMF from the large scale structure simulation from Ref.~\cite{ryu,miniati}
has so far been implemented in CRPropa, but any magnetic field model
can be used. Within the public package CRPropa, only a small subgrid
of the simulations from~\cite{ryu,miniati} is provided in order to
allow simple tests. Fig.~\ref{fig4} shows a 2D cross section through
the environment of a galaxy cluster from this simulation. In
this simulation, the magnetic fields follow the baryon density, and in
particular the regions that are filled with sub-$\mu$G fields are
quite extended around the large-scale structures (with a typical extension of
a few Mpc). This is due, in particular, to the fact that magnetic
fields are generated at the LSS shocks within that model. Of course,
the properties of $\gamma$-ray sources associated with UHECR sources as
well as the feasibility of ``charged particle astronomy'' depend
strongly on the magnetic field model~\cite{Sigl:2004gi},~\cite{dolag}.

Large scale structure simulations usually cover only a small fraction of
today's Universe, typically of order 100 Mpc in linear scale. Since sources
at much larger, cosmological distances can contribute to the fluxes of UHECR
below the GZK cutoff, of photons below $\sim\,$TeV and of neutrinos, the EGMF
and source distributions are periodically continued in the 3D version of the
code.
EGMF with homogeneous statistical properties and power law spectra in
Fourier space (e.g a Kolmogorov spectrum) have also been implemented
in the package.

\section{Simple Applications}

We present here applications of CRPropa that are obtained with
very simple configurations requiring little CPU time. The results
can easily be compared with previous results from the literature.

\begin{figure}
\begin{center}
\includegraphics[width=0.8\textwidth]{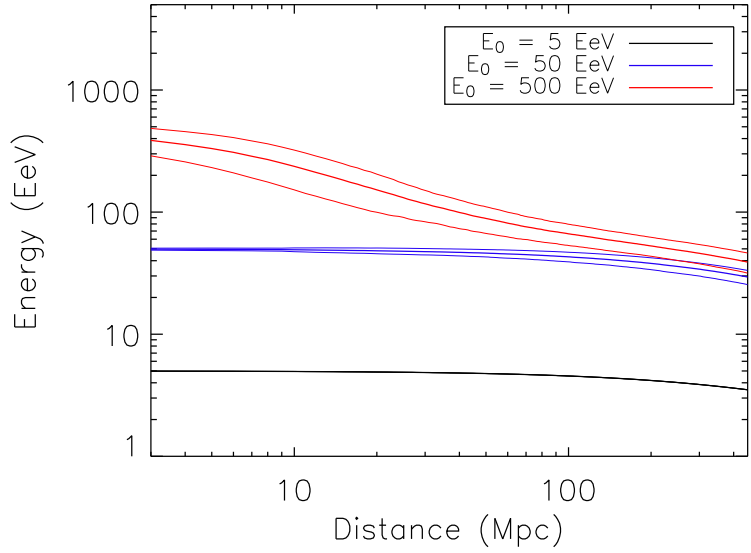}
\caption{\label{traj1d}Evolution of the energy of nucleons as a
function of propagation distance, for initial energies of 5, 50 or
500 EeV. The thin lines indicate the dispersion induced by the
stochasticity of pion production.}
\end{center}
\end{figure}

Fig.~\ref{traj1d} shows the averages and dispersions of the energy of
nucleons in a one-dimensional simulation, as a function of propagated
distance for various initial energies.
Using SOPHIA automatically enables us to reproduce the
stochasticity of pion production.

\begin{figure}
\begin{center}
\includegraphics[width=0.8\textwidth]{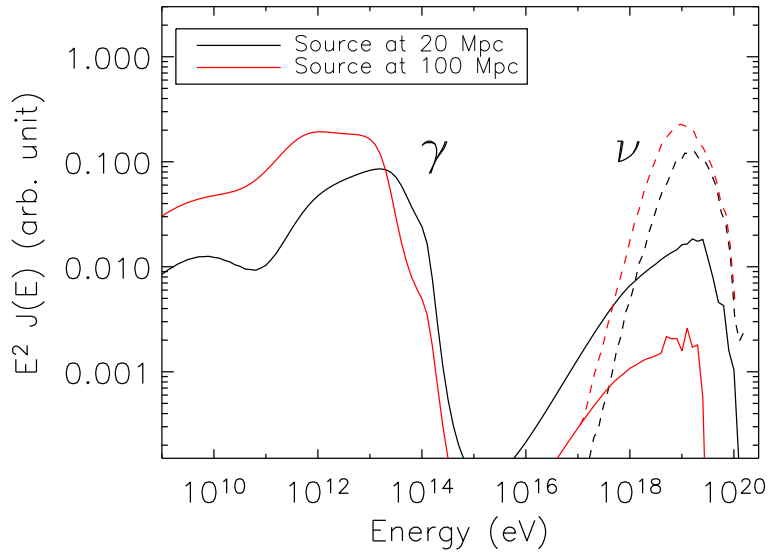}
\caption{\label{secondaries}Spectrum of secondary photons and
neutrinos (all flavors added) generated by pion and pair production 
from a single UHECR source at a given distance. We consider here a 
one-dimensionnal model, with an injection
spectral index $\alpha = 2$ for the UHECRs. A uniform magnetic field
of 0.1 nG is assumed. Note that below $\sim$TeV the $\gamma-$ray
flux would be spread over several degrees and that, as shown in
Fig.~\ref{fig:cascade_cut}, the 1D approximation of the EM cascade
does not significantly affect the accuracy of the $\gamma-$ray flux
arriving within such angles.
The fluctuations at the highest energies are statistical.}
\end{center}
\end{figure}

Fig.~\ref{secondaries} shows the spectra of secondaries generated
during the one-dimensional propagation of UHECRs from a source
located at 20 Mpc or 100 Mpc from the observer. Note that the
neutrino flux increases with distance to the source, whereas the
photon flux above $\sim10^{14}\,$eV decreases, but the photon flux
below this energy increases. This is because more secondary neutrinos
and EM particles are produced for larger propagation distances, but
EM particles above $\sim10^{14}\,$eV are quickly degraded and cascade
down to sub-PeV energies. A more detailed analysis of
the fluxes of secondaries from a single UHECR source (e.g. the
relative contribution of pair production and pion production on the
$\gamma$-ray flux) can be found in Ref.~\cite{Armengaud:2005cr}. The
study of secondary photons from UHECR sources has also been carried
out in various situations in
Refs~\cite{Gabici:2005gd,Ferrigno:2004am,Rordorf:2004jp,Inoue:2005vz,aharonian}.

\begin{figure}
\begin{center}
\includegraphics[width=0.8\textwidth]{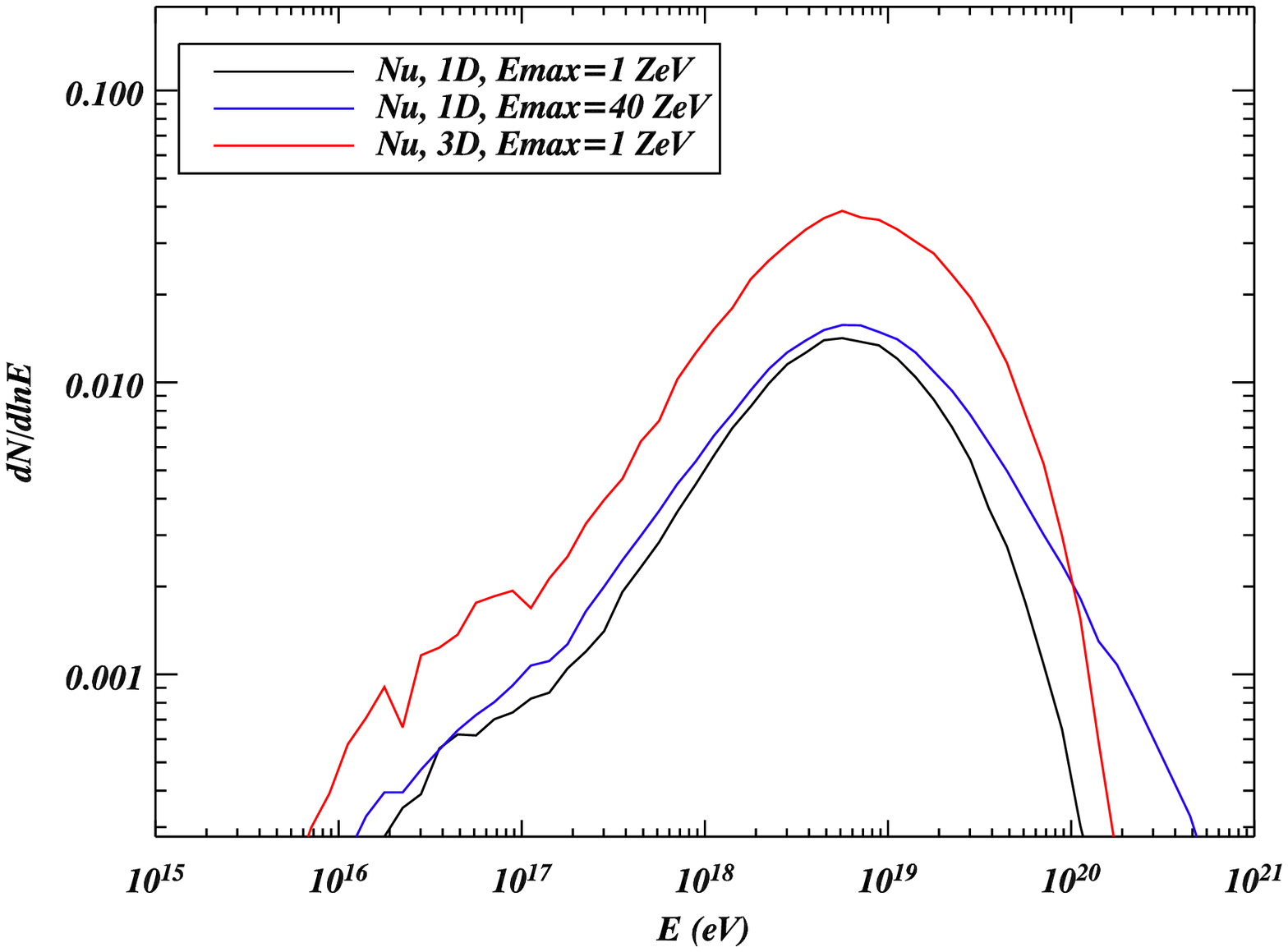}
\caption{\label{sourcenu} Secondary neutrinos (all flavors added) 
from a nearby source of
UHECRs with a given luminosity. The flux increases at high energies
both with maximum UHECR acceleration energy and with the strength of
magnetic fields surrounding the source. The
fluctuations at low energy are statistical. The $y$-axis is in
arbitrary units.}
\end{center}
\end{figure}

Fig.~\ref{sourcenu} shows the spectra of secondary neutrinos from a
source located at 20 Mpc from an observer, depending in particular on
the magnetic field effects. It is remarkable that, for a given source
luminosity, the flux of secondary neutrinos is increased by a factor
of more than two due to the enhancement of the UHECR propagation
distance generated by the $\mu$G-level magnetic fields that surround
this source.

\begin{figure}
\begin{center}
\includegraphics[width=0.8\textwidth]{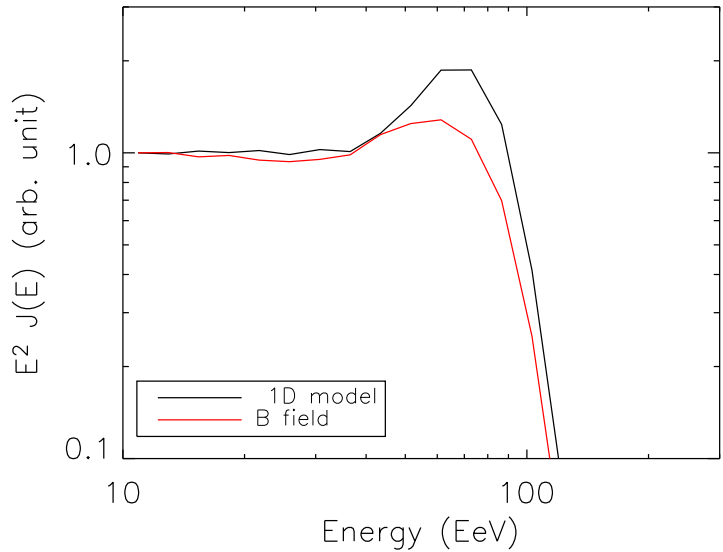}
\caption{\label{source_3d}UHECR spectrum from a source located at 100
Mpc from an observer, injecting protons with a spectrum $\propto E^{-2}$
up to $E_{\rm max}=10^{21}$ eV. The red curve
is obtained from a full 3-dimensional simulation, where the source is
embedded in a region with $\mu$G fields over a few Mpc.}
\end{center}
\end{figure}

Fig.~\ref{source_3d} compares the spectral shape of UHECRs
from a source located at 100 Mpc from an observer, depending on the
presence of magnetic fields around the source. If magnetic fields of amplitude 
$\sim \mu G$ surround the source over a few Mpc, the observed spectrum
is clearly modified: 1) there is a dispersion in the true propagation
distance, compared to a fixed propagation distance of 100 Mpc. 
This reduces the amplitude of the "bump"; 2) the mean propagation
distance is increased compared to 100 Mpc. This leads to a GZK
cut-off at slightly lower energies.

\begin{figure}
\begin{center}
\includegraphics[width=0.8\textwidth]{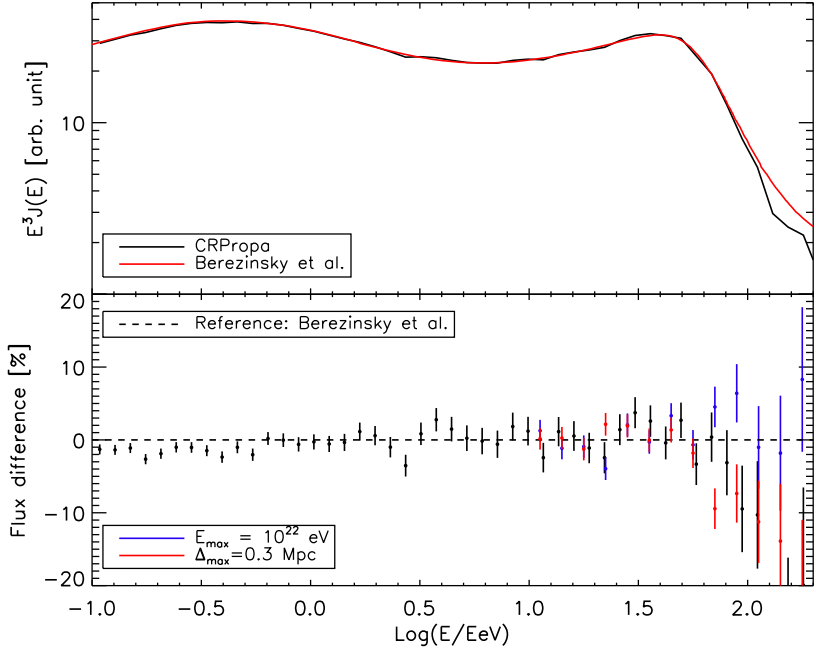}
\caption{\label{test_pairprod_dip} Top: Comparison of the spectra obtained with CRPropa to the one found in~\cite{Berezinsky:2002nc} for a model of cosmologically distributed sources. The specific parameters of the model correspond to the red curve of Fig.14 in~\cite{Berezinsky:2002nc}. We use a propagation stepsize of 1Mpc and a maximum injection energy of $10^{21}$ eV. Bottom: Relative difference with respect to the curve of~\cite{Berezinsky:2002nc}(black). The error bars are the statistical uncertainties due to the finite number of propagated nucleons. Red: same, for a simulation using a stepsize of 0.3 Mpc. Blue: same, for a simulation using a maximal injection energy of $10^{22}$ eV.}
\end{center}
\end{figure}

Fig.~\ref{test_pairprod_dip} compares the spectra obtained with CRPropa to the one presented on the red curve of Fig.14 in~\cite{Berezinsky:2002nc} for a model of cosmologically distributed proton sources with spectral index $\gamma_s=2.6$ and a source evolution parameter $m=2.4$. We see that, for a given model, the spectrum estimations obtained with our Monte-Carlo method and with a direct integration of the transport equations (for~\cite{Berezinsky:2002nc}) agree within a few \%.
The blue and red curves of the lower panel in Fig.~\ref{test_pairprod_dip} show the influence of two numerical parameters on the accuracy of the derived spectrum at the highest energies. The maximum injection energy allowed in the Monte-Carlo has an influence at energies above $10^{20}$ eV, in agreement with results shown, for example, in Fig.5 of~\cite{Berezinsky:2002nc}. The use of a propagation stepsize of 0.3 Mpc instead of 1 Mpc does not lead to a significant change in the simulated spectrum. Other tests showed that using a propagation stepsize of 5 Mpc instead of 1Mpc results in a $\sim 10\%$ overestimation of the spectrum in the specific energy range $100 < E < 160$ EeV, and an underestimation of the spectrum at higher energies. A 1 Mpc stepsize is therefore reasonable to reach the typical accuracies required for comparison with current and forecoming experimental data.

\section{Conclusions}

We have presented the first public package to study systematically the
properties of the propagation of UHECRs and their secondaries in a
structured magnetized Universe. We have
detailed the interactions that are already implemented, and presented a
few simple examples obtained directly by running the CRPropa code.

A major advantage of CRPropa is its large modularity, which should
allow various users to implement their own modules, adapted to
specific UHECR propagation models. Many possible upgrades of the
CRPropa package can be considered: This includes the implementation
of non-uniform grids for magnetic field models, of UHE nuclei and secondary
neutrinos and EM particles from their interactions, of inhomogeneous
low energy target photon backgrounds for the UHE nuclei and EM cascade
interactions, and of hadronic interactions with the baryon gas in
dense parts of the large scale structure. Finally, interactions of UHE
neutrinos with relic neutrinos of arbitrary mass and clustering properties
could also be implemented, including the resulting secondary particles.

\ack
FM acknowledges partial support by the Swiss Institute of Technology through a Zwicky
Prize Fellowship. We thank all the people who built the previous
codes from which the development of CRPropa has largely taken profit,
in particular
Martin Lemoine, Gianfranco Bertone, Claudia Isola, and Sangiin Lee. We also thank S\'ebastien Renaux-P\'etel for useful tests.

CRPropa makes use of the public code SOPHIA~\cite{sophia}, and the
TinyXML~\cite{tinyxml}, CFITSIO~\cite{cfitsio} and
CLHEP~\cite{clhep} libraries.

\end{document}